\documentclass[twocolumn]{aastex701}

\graphicspath{{./figures/}}
\makeatletter
\def\input@path{{tables/}}
\makeatother

\usepackage{xspace}

\newcommand{\bjdtdb}{\ensuremath{\mathrm{BJD}_\mathrm{TDB}}\xspace}
\newcommand{\bjd}{\ensuremath{\mathrm{BJD}}\xspace}

\newcommand{\rhostar}{\ensuremath{\rho_{\star}}\xspace}
\newcommand{\Rjup}{\ensuremath{R_\mathrm{J}}\xspace} 
\newcommand{\Mjup}{\ensuremath{M_\mathrm{J}}\xspace}
 
\newcommand{\Mearth}{\ensuremath{M_\oplus}\xspace}

\newcommand{\Mp}{\ensuremath{M_p}\xspace}
\newcommand{\rhop}{\ensuremath{\rho_p}\xspace}

\newcommand{\Teq}{\ensuremath{T_\mathrm{eq}}\xspace}

\newcommand{\ecosw}{\ensuremath{e\cos{\omega}}\xspace}
\newcommand{\esinw}{\ensuremath{e\sin{\omega}}\xspace}
\newcommand{\secosw}{\ensuremath{\sqrt{e}\cos{\omega}}\xspace}
\newcommand{\sesinw}{\ensuremath{\sqrt{e}\sin{\omega}}\xspace}
\newcommand{\gammadot}{\ensuremath{\dot{\gamma}}\xspace}
\newcommand{\Qprime}{\ensuremath{Q_p^\prime}\xspace}
\newcommand{\ms}{\ensuremath{\mathrm{m}\,\mathrm{s}^{-1}}\xspace}

\newcommand{\msyr}{\ensuremath{\mathrm{m}\,\mathrm{s}^{-1}\,\mathrm{yr}^{-1}}\xspace}

\newcommand{\gcc}{\ensuremath{\mathrm{g}\,\mathrm{cm}^{-3}}\xspace}
\newcommand{\ergs}{\ensuremath{\mathrm{erg}\,\mathrm{s}^{-1}}\xspace}

\begin{document}

\title{``Popcorn Planets'' are Not Actively Inflated by Eccentricity Tides}

\author[0000-0001-7961-3907]{Samuel~W.~Yee}
\altaffiliation{51 Pegasi b Fellow}
\affiliation{Center for Astrophysics $\vert$ Harvard \& Smithsonian, 60 Garden Street, Cambridge, MA 02138, USA}
\affiliation{Department of Physics \& Astronomy, University of California Los Angeles, Los Angeles, CA 90095, USA}
\email{syee@astro.ucla.edu}

\author[0000-0003-2527-1475]{Shreyas~Vissapragada} 
\affiliation{Carnegie Science Observatories, 813 Santa Barbara Street, Pasadena, CA 91101, USA}
\email{svissapragada@carnegiescience.edu}

\begin{abstract}
Recent discoveries have revealed a population of ``popcorn planets'' that have masses similar to that of Neptune but radii comparable to Jupiter, leading to exceptionally low bulk densities $\rhop \lesssim 0.3$\,\gcc.
Their anomalously-inflated radii, along with recent JWST atmospheric observations, suggest a source of internal heating.
Because these planets are nominally too cool to be affected by the hot Jupiter inflation mechanism, dissipation of eccentricity tides within the planet has been proposed as a leading explanation for the source of this heat flux. Using the MAROON-X spectrograph on Gemini-North, we conducted a high-precision radial-velocity campaign to precisely measure the eccentricities of three of these popcorn planets: WASP-107\,b, TOI-1173\,b, and HAT-P-18\,b.
We constrained their eccentricities below $e < 0.03$--$0.05$ to 95\% confidence, decisively ruling out active heating from eccentricity tides as the cause of these planets' inflated radii (except for the unlikely scenario in which their tidal quality factors are less than the Earth's). An alternative heating mechanism is likely responsible for inflating these planets. Our measurements also provide new constraints on $e\cos\omega$, significantly shrinking the eclipse timing uncertainties to better than $\pm2.5$~hr and allowing for confident scheduling of thermal emission measurements for these enigmatic planets.
\end{abstract}


\section{Introduction}

The discovery of the first transiting extrasolar giant planet \citep{Charbonneau2000,Henry2000} revealed that its radius was significantly larger than expected based on internal structure models \citep[e.g.,][]{Bodenheimer2001,Guillot2002}.
The mechanism behind this ``hot Jupiter radius anomaly'' has not yet been fully identified \citep[see e.g., review by][]{Thorngren2024}, but the level of radius inflation is strongly correlated with incident flux \citep{Laughlin2011, Weiss2013,Thorngren2018}, becoming negligible for planets with equilibrium temperatures~$\Teq < 1000$~K.

Recently, a number of planets have been discovered that have highly inflated radii, but which are too cool to be affected by the same inflation mechanism.
These have masses similar to that of Neptune $0.05\,\Mjup < \Mp < 0.2\,\Mjup$ but radii similar to that of Jupiter, resulting in planetary bulk densities of $\rhop \lesssim 0.3\,\gcc$.
These cool, puffy Neptunes, which can be thought of as higher-mass analogues to the ``super-puffs'' \citep{Lee2016}, include WASP-107\,b \citep{Anderson2011}, HAT-P-18\,b \citep{Hartman2011}, and TOI-1173A\,b \citep{YanaGalarza2024a,Polanski2024}.

The exceptionally low densities of these ``popcorn planets'' are not easily explained.
One possibility is that these planets have low core masses of $M_\mathrm{core}\lesssim 5\,\Mearth$ and large gas envelope mass fractions \citep{Lee2019, Piaulet2021}.
This goes against predictions from core accretion theory \citep[e.g.,][]{Pollack1996}, which posits that larger cores $M_\mathrm{core} \gtrsim 10\,\Mearth$ are required to initiate runaway accretion of gas envelopes much more massive than the original core.
Another explanation is that these planets have more Neptune-like gas envelope mass fractions ($f_\mathrm{env} \sim 10$\textendash$30\%$), but are inflated by a source of excess heat within their interiors \citep{Millholland2020}.
Recent \textit{JWST} observations of WASP-107\,b \citep{Dyrek2024, Welbanks2024,Sing2024} and HAT-P-18\,b \citep{Fu2022, FournierTondreau2024} have found evidence for such internal heating. CH$_4$ is expected in equilibrium for these relatively cool planets, but is observed to be depleted, plausibly due to vertical mixing from the planet's anomalously-heated interior \citep{Fortney2020}. However, such observations do not identify the source of the excess heat.

Currently, a leading hypothesis is that the internal heating results from dissipation of eccentricity tides \citep[e.g.,][]{Millholland2020}.
The most well-studied example of this population, WASP-107\,b, was previously reported to have a potentially non-zero eccentricity of $e = 0.06\pm0.04$ \citep{Piaulet2021}.
Ongoing circularization of the planet's orbit due to dissipation of tidal energy could result in significant energy deposited within the planet \citep{Leconte2010}, inflating its atmosphere.
On the other hand, the rapid tidal dissipation necessary to explain the radius inflation would also imply fast orbital circularization, so for the orbit to remain eccentric given the $\sim$Gyr ages of the stars, the planets must have either recently arrived on their current orbits via dynamical migration \citep[e.g.,][]{Yu2024}, or an external perturber must be actively driving their eccentricities \citep[but see][]{Mardling2007, Batygin2025}.

There is a straightforward test of this idea: if eccentricity tides really are heating these planets, then their orbits should be measurably eccentric. While the previous measurement of WASP-107\,b's eccentricity permits modest values, it is also consistent with a circular orbit within 1.5$\sigma$, insufficient to claim a nonzero eccentricity especially when considering the \citet{LucySweeney71} bias.
The other popcorn planets are less well-studied, and have even poorer constraints on their eccentricities.

This Letter presents new precise radial-velocity (RV) measurements of three stars hosting popcorn planets: WASP-107, TOI-1173, and HAT-P-18. Our goal was to determine whether these planets reside on eccentric orbits or not. We place new stringent constraints of $e \lesssim 0.05$ for all three planets, ruling out eccentricity tides as the source of the internal heat flux given plausible tidal quality factors for the planets.

\section{Observations \& Analysis} \label{sec:observations}

\begin{deluxetable*}{lccccc}
\tablecaption{New MAROON-X RV Measurements \label{tab:mx_rvdata}}
\tablehead{
\colhead{Target} & \colhead{Instrument} & \colhead{\bjdtdb} & \colhead{RV [\ms]} & \colhead{$\sigma_\mathrm{RV}$ [\ms]} & \colhead{$T_\mathrm{exp}$ [s]}
}
\startdata
TOI-1173 & MX-Blue & 2460692.170633 & 4.15 & 0.78 & 900  \\
TOI-1173 & MX-Blue & 2460694.174731 & 5.16 & 1.25 & 900  \\
TOI-1173 & MX-Blue & 2460695.134909 & -3.29 & 0.89 & 900  \\
TOI-1173 & MX-Blue & 2460696.158947 & -10.46 & 0.57 & 900  \\
TOI-1173 & MX-Blue & 2460759.077697 & -6.54 & 0.55 & 900%
\enddata
\tablecomments{This table is available in its entirety in machine-readable form in the online article.}
\end{deluxetable*}
\begin{deluxetable*}{cccccc}
\tablecaption{Best-fit Eccentricities from RV and Transit Fitting \label{tab:ecc_summary}}
\tablehead{
\colhead{Planet} & \colhead{\secosw} & \colhead{\sesinw} & \multicolumn{3}{c}{Eccentricity upper limits} \\
& & & $68.3\%$ & $95.5\%$ & $99.7\%$ 
}
\startdata
TOI-1173 b & $-0.053^{+0.053}_{-0.065}$ & $0.033^{+0.099}_{-0.095}$ & $<0.018$ & $<0.037$ & $<0.062$\\
WASP-107 b & $0.077^{+0.095}_{-0.065}$ & $0.02^{+0.12}_{-0.11}$ & $<0.028$ & $<0.052$ & $<0.081$\\
HAT-P-18 b & $-0.046^{+0.057}_{-0.061}$ & $0.029^{+0.096}_{-0.081}$ & $<0.015$ & $<0.030$ & $<0.048$
\enddata
\end{deluxetable*}

We observed TOI-1173, WASP-107 and HAT-P-18 with the MAROON-X echelle spectrograph \citep{MAROONX_Seifahrt2016,MAROONX_Seifahrt2018,MAROONX_Seifahrt2020,MAROONX_Seifahrt2022} on the Gemini-North telescope (Program IDs 2025A-GN-Q-110, 2025B-GN-Q-204).
We designed the high cadence, high precision observational campaign to precisely and accurately measure the planets' eccentricities.
We obtained 30 new observations of TOI-1173, 20 of WASP-107, and 24 of HAT-P-18, with exposure times of 900~s, 600~s, and 600~s respectively.
The MAROON-X data were reduced and 1-dimensional spectra extracted with the standard instrument pipeline.\footnote{\url{https://www.gemini.edu/instrumentation/maroon-x/data-reduction}}
Wavelength calibration of the MAROON-X spectra is performed via simultaneous observations of a Fabry-P\'{e}rot etalon.

We measured precise RVs from the reduced spectra using a modified version of the \texttt{SERVAL} package \citep{SERVAL_Zechmeister2018}.
In brief, \texttt{SERVAL} constructs an empirical high signal-to-noise ratio (SNR) spectral template by co-adding each observation after shifting them onto a reference wavelength scale.
Relative RVs between each individual spectrum and the high SNR template are then computed via least-squares fitting on an order-by-order basis.
MAROON-X has independent ``red'' and ``blue'' channels with separate cross-dispersers and detectors; we treated the two channels separately and computed a weighted mean of the RVs across all orders to arrive at independent final RVs for each channel.
The median per-point instrumental uncertainties achieved were 1.2~\ms and 0.66~\ms for TOI-1173 in the red and blue channels respectively, 1.4~\ms and 0.94~\ms for WASP-107, and 2.4~\ms and 1.5~\ms for HAT-P-18.
The RV measurements are provided in Table \ref{tab:mx_rvdata}.

We jointly fitted the new MAROON-X RV measurements, previously-published RV data, and transit light curves from NASA's Transiting Exoplanet Survey Satellite (TESS) mission \citep{TESS_Ricker2014}, the latter of which provide additional constraints on the planets' orbital ephemerides and \esinw component of the eccentricity vector (see Appendix \ref{sec:fitting_details} for further details).
These data were modeled with the \texttt{exoplanet} package \citep{Exoplanet_Joss} using a standard Keplerian model for the RVs and a quadratic limb-darkened transit model \citep{Exoplanet_Agol20} for the photometry.
We fitted for an independent offset and jitter term to account for any underestimated instrumental uncertainties for each RV instrument.
RVs from the two arms of MAROON-X were treated separately, as were different MAROON-X observing runs, allowing for potential $\sim\ms$ offsets between runs \citep{MAROONXOffsets_Basant2025}.

The planet eccentricities were parameterized in terms of \secosw and \sesinw, which imposes an implicit uniform prior on $e$ \citep{Anderson2011}.
For the WASP-107 system, we fitted the RVs with a two-Keplerian model to account for the known second planet, the wide-orbiting ($P \approx 1000$~days) giant planet WASP-107\,c \citep{Piaulet2021}.
For TOI-1173 and HAT-P-18, we fitted for a linear RV slope \gammadot to allow for acceleration due to any unseen perturbers, although in both cases the inferred \gammadot was within 2$\sigma$ of zero.

\begin{figure}
    \epsscale{1.1}
    \plotone{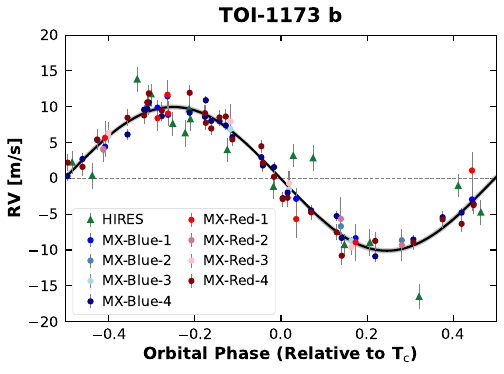}
    \plotone{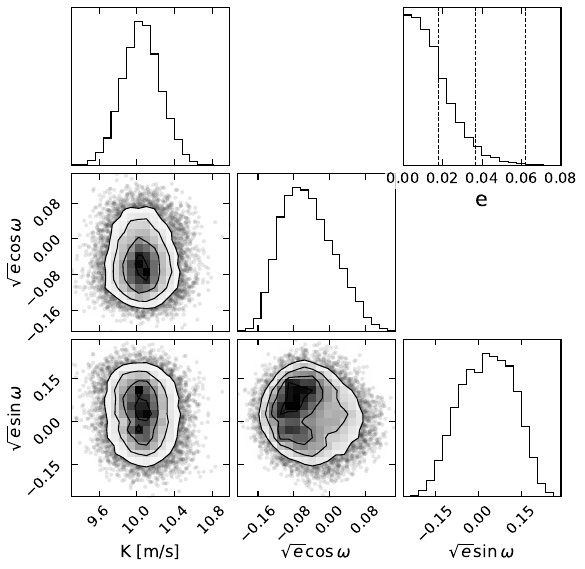}
    \caption{\textbf{Top:} RV measurements of TOI-1173, phase-folded to the ephemeris for TOI-1173\,b. Per-instrument offsets as well as a global RV trend have been subtracted from the data. The error bars represent the quadrature sum of the reported instrumental uncertainties and an additional RV jitter term unique for each instrument.
    The black line shows the maximum a-posteriori model for the RV variation caused by TOI-1173\,b, while the 1-$\sigma$ limits are shown in light gray.
    \textbf{Bottom:} Corner plot showing the posterior distribution for the key RV fitted parameters $K$, \secosw, and \sesinw. The inset panel shows the posterior distribution for the planet's orbital eccentricity $e$, with vertical dashed lines denoting the 68\%, 95\%, and 99.7\% upper limits.}
    \label{fig:toi1173}
\end{figure}

\begin{figure}
    \epsscale{1.1}
    \plotone{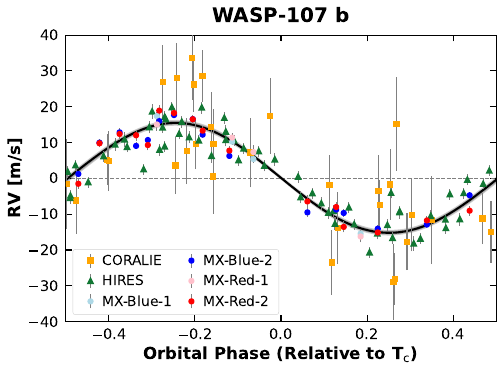}
    \plotone{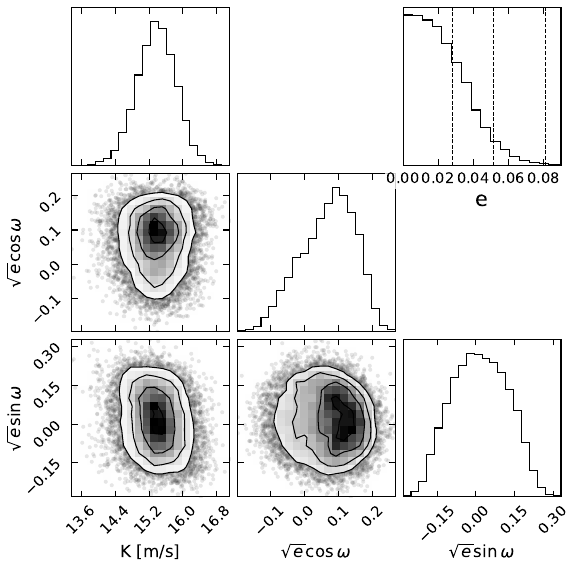}
    \caption{Same as Figure \ref{fig:toi1173}, but for WASP-107. In the top panel, the RV variation due to the outer planet WASP-107\,c has been subtracted from the data.}
    \label{fig:wasp107}
\end{figure}

\begin{figure}
    \epsscale{1.1}
    \plotone{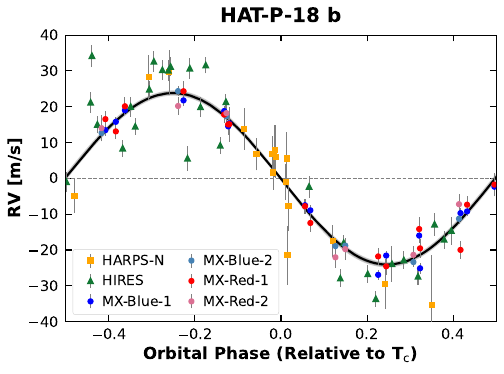}
    \plotone{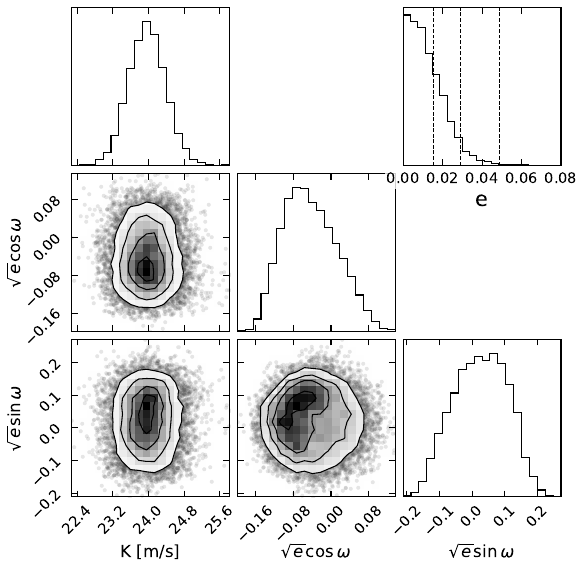}
    \caption{Same as Figure \ref{fig:toi1173}, but for HAT-P-18.}
    \label{fig:hatp18}
\end{figure}

We estimated the posterior probability distributions of the fitted parameters using a No U-Turn Sampler (NUTS; \citealp{NUTS_Hoffman}), an adaptive Hamilton Monte Carlo (HMC) sampler as implemented in \texttt{pymc} \citep{PyMC}.
We ran four parallel HMC chains and assessed convergence by ensuring the rank-normalized $\widehat{R}$ statistic is below 1.001 \citep{Rhat_Vehtari2021} and an effective sample size of $> 10{,}000$ samples.
The phase-folded RV data and best-fit models are plotted in Figures \ref{fig:toi1173}, \ref{fig:wasp107}, and \ref{fig:hatp18}, along with corner plots showing the posterior distributions of $K$, $\sqrt{e}\cos{\omega}$, and $\sqrt{e}\sin{\omega}$.
Table \ref{tab:ecc_summary} provides the new eccentricity constraints for each planet while the full set of fitted and derived physical parameters are provided in Tables \ref{tab:toi1173_params}--\ref{tab:hatp18_params} in Appendix \ref{sec:fitting_details}.

\section{Popcorn Planets are Not Actively Inflated by Eccentricity Tides}

Our analysis of the new, precise RV observations indicates that the popcorn planets TOI-1173\,b, WASP-107\,b, and HAT-P-18\,b do not have measurably eccentric orbits.
As Figures \ref{fig:toi1173}, \ref{fig:wasp107}, and \ref{fig:hatp18} show, the components of the vector (\secosw, \sesinw) are consistent with zero, allowing us to place strong upper limits on the true eccentricities of the planets (Table \ref{tab:ecc_summary}).
The 95\% upper limits are $e < 0.037$ for TOI-1173\,b, $e < 0.052$ for WASP-107\,b, and $< 0.030$ for HAT-P-18\,b.
Previously published eccentricity estimates were also consistent with zero but permitted larger eccentricities due to the lower precision and smaller quantity of data analyzed.
For example, \citet{Piaulet2021} reported $e = 0.06\pm0.04$ for WASP-107\,b.
Our new data allow us to rule out $e \geq 0.06$ for WASP-107\,b with 97.7\% confidence.
We also performed robustness checks, described in Appendix \ref{sec:robustness}, which demonstrate that our eccentricity estimates are biased by no more than $b_e = 0.02$, smaller than our derived upper limits.

The low eccentricities of these popcorn planets imply that their inflated radii cannot be the result of heating due to dissipation of eccentricity tides within the planet.
Assuming spin-orbit pseudo-synchronization, the tidal luminosity due to dissipation in the planet is given by \citep[e.g.][]{Leconte2010}:
\begin{equation}
L_\mathrm{tide} = \frac{21}{2}\frac{1}{Q_p^\prime} \frac{G M_\star^2}{R_p}\left(\frac{R_p}{a}\right)^6 ne^2,
\end{equation}
where $Q_p^\prime \equiv Q/k_2$ is the modified tidal quality factor.
While the precise nature of tidal dissipation within planets is not fully understood, empirical constraints based on population studies of tidal circularization of hot Jupiters suggest $\Qprime = 10^5$\textendash$10^6$ \citep[e.g.,][]{Quinn2014,Bonomo2017,Mahmud2023}, while $\Qprime=10^5$ has been reported for Jupiter \citep{Lainey2016}.
If we assume $\Qprime=10^5$ and the 68\% upper limit on eccentricity derived for each planet, this implies tidal luminosities $< 4\times10^{23}\,\ergs$, $6\times10^{24}\,\ergs$ and $2\times10^{24}\,\ergs$ for TOI-1173\,b, WASP-107\,b, and HAT-P-18\,b ($L_\mathrm{tide} / L_\mathrm{irr} < 2\times 10^{-5}$, $6\times10^{-4}$, $1\times10^{-4}$), far too small for significant radius inflation \citep{Thorngren2018, Piaulet2021, Batygin2025}.

\citet{Millholland2020} suggested that if planets like WASP-107\,b are inflated by tidal heating, their bulk properties could be consistent with Neptune-like envelope mass fractions of $f_\mathrm{env}\sim10$--30\%, rather than the $\sim $60--80\% required by interior structure models to match the observed planet density without internal heating \citep{Piaulet2021,Sing2024}.
We used the models of \citet{Millholland2019,Millholland2020} to estimate joint constraints on $f_\mathrm{env}$ and \Qprime given our new eccentricity measurements.
We find that $\Qprime < 10^3$ would be required to inflate a planet with $f_\mathrm{env} = 30\%$ to the radius of WASP-107\,b.
We note that this constraint is roughly two orders of magnitude lower than that found by \citet{Millholland2020}, who used $e = 0.13$ for their calculations, as expected given the scaling of tidal luminosity as $\propto e^2$.
If we take a Neptune-like value for $k_2 \approx 0.15$ \citep{Lainey2016}, this would imply a tidal quality factor $Q \lesssim 100$, a value similar to that inferred for the rocky planets in our solar system and orders of magnitude lower than that for the giant planets.

Furthermore, the rapid tidal dissipation implied by such a small tidal quality factor is difficult to reconcile with the old age of these systems, since the orbits should have quickly circularized.
This was pointed out by \citet{Batygin2025}, who additionally showed that WASP-107\,b's eccentricity cannot be maintained by secular interactions with the outer planet in the system.
Using the equations of \citet{Leconte2010}, we find that for $\Qprime = 10^3$, WASP-107\,b's orbit would have circularized from $e = 0.9$ to $e < 0.028$ within just a few Myr, and the circularization timescales for the other planets are similar.
If indeed these planets are actively undergoing such rapid circularization, we must be observing them during a special, short-lived phase for them to still have sufficient remnant eccentricity to cause significant inflation.
One caveat is that the response of the planet's radius to changes in internal luminosity may not be instantaneous --- for example, \citet{Thorngren2021} inferred a ``deflation timescale'' of up to 0.5~Gyr for hot Jupiters, which could extend the window during which a planet may stay inflated even after the conclusion of orbit circularization.

\begin{figure}
    \epsscale{1.2}
    \plotone{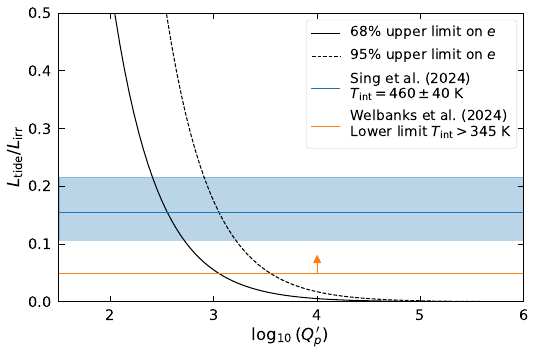}
    \caption{Luminosity due to eccentricity tides for WASP-107\,b, as a function of \Qprime. The solid and dashed lines correspond to the 68\% and 95\% upper limits on the tidal luminsoity.
    Based on \textit{JWST} atmospheric observations, \citet{Sing2024} measured an internal temperature of $460\pm40$~K (blue shaded region). For eccentricity tides to be the source of this internal heat flux, the tidal quality factor would have to be $\log_{10}{\Qprime} \lesssim 2.5$ given our 1-$\sigma$ upper limit on $e$.}
    \label{fig:wasp107_tidal_luminosity}
\end{figure}

We can also make a comparison to the luminosities implied by the observed internal heat flux, as inferred from \textit{JWST} observations of CH$_4$ depletion \citep{Fortney2020} for WASP-107\,b and HAT-P-18\,b.
In the case of WASP-107\,b, \citet{Sing2024} reported an internal temperature of $T_\mathrm{int} = 460\pm40$~K, while \citet{Welbanks2024} placed a 3-$\sigma$ lower limit of $T_\mathrm{int} > 345$~K.
Figure \ref{fig:wasp107_tidal_luminosity} shows that if dissipation of eccentricity tides is solely responsible for this internal heat flux, it would require $\log_{10}{\Qprime} \lesssim 2.5$ for WASP-107\,b, again much lower than expected for gaseous planets.
Circularization of the current eccentricity given a Jupiter-like $Q^\prime \sim 10^5$ would produce a heat flux more than two orders of magnitude too small to explain the inferred values from the atmospheric observations.
\citet{Batygin2025} used similar arguments to argue against eccentricity tides for WASP-107\,b, but our new constraints are even more stringent thanks to our updated eccentricity measurements.

These results indicate that the popcorn planets are not being inflated by ongoing dissipation of eccentricity tides in their interiors, unless unphysically low tidal quality factors are invoked.
Still, their low densities could be reconciled with predictions from core accretion theory, as well as their apparent high internal heat fluxes, if there is an alternative source of heating.
\citet{Millholland2020} demonstrated that obliquity tides, arising from the angle between the planet's orbital and spin axes (the planetary obliquity), could be a significant contributor to tidal heating.
This remains a viable explanation if the planet has a large obliquity --- for WASP-107\,b, the tidal luminosity due to obliquity tides is $\sim 10^{27}\,\ergs$ for $\epsilon \gtrsim 40^\circ$.

Meanwhile, \citet{Batygin2025} proposed that Ohmic dissipation could reproduce the observed internal heat flux for WASP-107\,b, depending on specific assumptions about atmospheric ionization, wind speeds, and the planet's magnetic field strength.
A test for these and other proposed mechanisms would be to explain why such significant radius inflation is observed for planets like WASP-107\,b and TOI-1173\,b, but does not operate in other planets with a similar mass and temperature. 

Given the intriguing thermal states of popcorn planets, secondary eclipse observations are likely to be a fruitful avenue for future investigations. Internal heating changes the gradient of the $T$-$P$ profile at pressures that are accessible in eclipse \citep{Guillot2010, Fortney2020}, so emission spectroscopy could more directly probe the anomalous heating mechanism. These measurements have been challenging to schedule in the past: eclipses of eccentric planets are offset by $\frac{2P}{\pi}e\cos\omega$ from orbital phase 0.5 \citep{Winn2010}, and previous $e\cos\omega$ uncertainties have resulted in eclipse timing windows spanning $>12$~hr. Our observations have collapsed these uncertainties by factors of 3--5, resulting in 95\% confidence intervals of $\pm1.7$, $\pm2.4$, and $\pm1.2$~hr on the eclipse times of TOI-1173b, WASP-107b, and HAT-P-18b respectively. 

\section{Conclusion}
Popcorn planets have exceptionally low densities not easily reconciled with predictions from core accretion theory in the absence of anomalous heating driving radius inflation.
The upper atmospheres of two such planets are also observed to be depleted in CH$_4$, another sign of significant internal heating. 
Eccentricity tides are usually invoked to explain their hot interiors, but here we expose a fatal flaw for this argument: they do not reside on very eccentric orbits. With 23~hr of new MAROON-X data, we have constrained $e < 0.05$ to better than 95\% confidence for three archetypal popcorn planets: TOI-1173Ab, WASP-107b, and HAT-P-18b.

If the only thing powering radius inflation and CH$_4$ depletion in these systems is the active dissipation of eccentricity tides, then the planets would need unphysical tidal quality factors $\Qprime < 10^2$, less than that inferred for solar system rocky planets.
Another mechanism, for instance obliquity tides \citep{Millholland2020} or Ohmic dissipation \citep{Batygin2025}, is probably responsible for heating the interiors of popcorn planets, although we cannot rule out the possibility that their inflated radii could be the product of delayed cooling from a past epoch of orbit circularization.

Popcorn planets are an interesting subsample of the exoplanet population with many unusual features. In addition to their low densities and high internal heat fluxes, a few have been observed to be losing mass \citep[e.g.,][]{Spake2018,Paragas2021, Vissapragada2024}, and at least some of them may be on polar orbits \citep{Dai2017,Rubenzahl2021,Esposito2014}, perhaps indicative of a dynamically violent formation history. 
Continued observations of these mysterious objects may yet reveal their nature and tie these different clues into a coherent story explaining both their origin and present-day structures.

\begin{acknowledgments}
The authors thank the MAROON-X science team for providing the reduced data products, as well as the Gemini queue observers for performing the observations.
S.W.Y. acknowledges support from the Heising-Simons Foundation.
The authors thank Daniel Thorngren for helpful conversations at the onset of this project.

Based on observations obtained at the international Gemini Observatory, a program of NSF NOIRLab, which is managed by the Association of Universities for Research in Astronomy (AURA) under a cooperative agreement with the U.S. National Science Foundation on behalf of the Gemini Observatory partnership: the U.S. National Science Foundation (United States), National Research Council (Canada), Agencia Nacional de Investigaci\'{o}n y Desarrollo (Chile), Ministerio de Ciencia, Tecnolog\'{i}a e Innovaci\'{o}n (Argentina), Minist\'{e}rio da Ci\^{e}ncia, Tecnologia, Inova\c{c}\~{o}es e Comunica\c{c}\~{o}es (Brazil), and Korea Astronomy and Space Science Institute (Republic of Korea).

This work was enabled by observations made from the Gemini North telescope, located within the Maunakea Science Reserve and adjacent to the summit of Maunakea. We are grateful for the privilege of observing the Universe from a place that is unique in both its astronomical quality and its cultural significance.

\end{acknowledgments}




\facilities{Gemini-N (MAROON-X), TESS}

\software{\texttt{astropy} \citep{Astropy13,Astropy18,Astropy2022},
          \texttt{exoplanet} \citep{Exoplanet_Joss,Exoplanet_Zenodo} and its dependencies
          \citep{Exoplanet_Kipping13, Exoplanet_Luger18, Exoplanet_Arviz, Exoplanet_Agol20},
          \texttt{PyMC5} \citep{PyMC},
          \texttt{lightkurve} \citep{Lightkurve18}
          }

\appendix

\section{Fitting Details and Results \label{sec:fitting_details}}
\subsection{Archival RV Data}
We fitted the new MAROON-X RV measurements together with previously-published RVs.
For TOI-1173, this comprises 18 RV measurements from Keck/HIRES published by \citet{Polanski2024} and 20 from Gemini-N/MAROON-X (10 independent epochs in the red and blue arms) published by \citet{YanaGalarza2024a}, which were taken with somewhat lower precision than our new observations.
In the case of WASP-107, we made use of 31 CORALIE observations reported by \citet{Anderson2017} and 60 Keck/HIRES RVs published by \citet{Piaulet2021}.
For HAT-P-18, we incorporated the 31 Keck/HIRES RVs published by \citet{Knutson2014} and 17 HARPS-N RVs from \citet{Bonomo2017}.

\subsection{TESS Transit Light Curves}

The TESS light curves were also used in the fit to provide more precise constraints on each planets' orbital period and time of conjunction than can be achieved from the RVs alone.
TESS observed TOI-1173 with 2-minute cadence in nine sectors (14, 15, 21, 22, 41, 47, 48, 74, and 75), WASP-107 at 30-minute cadence in Sector 10 and 2-minute cadence in Sector 91, and HAT-P-18 at 2-minute cadence in Sectors 25, 26, and 79. 
In each case, the TESS data spans multiple years, providing very tight constraints on the orbital periods.
We used the light curves produced by the SPOC pipeline \citep{TESS_SPOC_Jenkins2016} and flattened and clipped the light curve to include only the out-of-transit data within $\pm1$ transit duration of the planetary transit.

\subsection{Fitting Priors}
We used uninformative priors for most of the fitting parameters.
We imposed Gaussian priors on the $P_\mathrm{orb}$ and $T_c$ derived from pre-TESS data; these came from \citet{Dai2017} based on K2 data for WASP-107, and from \citet{Hartman2011} for HAT-P-18\,b.
TOI-1173\,b was discovered by TESS and thus had no previous ephemeris available, so we used uniform priors on $P_\mathrm{orb}$ and $T_c$.
Because the transit observables depend on the combination of the stellar density \rhostar and $\frac{1+e\sin{\omega}}{\sqrt{1-e^2}}$, we took care not to bias the inferred eccentricities by allowing the stellar densities to be free parameters in the fit with a wide uniform prior.
We used the $(q_1,q_2)$ parameterization for the quadratic limb-darkening parameters in the transit model from \citet{Exoplanet_Kipping13}.
We allowed for possible uncorrected dilution in the TESS light curves by allowing the flux baseline to vary separately in each sector, placing a Gaussian prior with 1\% width. In all cases the best-fit dilution corrections were consistent with zero.

\subsection{Full Fit Results}

\begin{figure}
    \centering
    \includegraphics[width=0.5\textwidth]{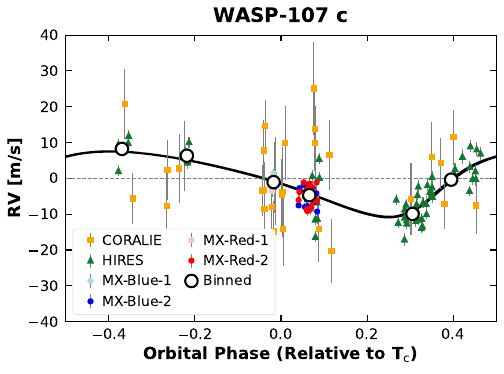}
    \caption{RV data for WASP-107, with the signal of the inner transiting planet removed and phase-folded to the period of WASP-107c. The best-fit model for WASP-107c is shown in black.}
    \label{fig:wasp107c_phased_rvs}
\end{figure}

We report the 68\% confidence intervals for each of the fitted and derived parameters for the three targets in Tables \ref{tab:toi1173_params}--\ref{tab:hatp18_params}.
Our results are within 2$\sigma$ of previously published parameters for these planets.
Apart from the eccentricities, our new RV measurements achieved notable improvements in the RV semi-amplitudes, which are now measured to better than 3\% for these three systems.
In Figure \ref{fig:wasp107c_phased_rvs}, we show our best-fitting model for the outer planet in the WASP-107 system, although because our RV campaign was designed to have a high observational cadence to isolate the orbital motion due to the inner planet, we did not obtain significant improvements for the orbital parameters of WASP-107c.

\begin{deluxetable}{lcc}
\tablecaption{TOI-1173 Fitted and Derived Parameters \label{tab:toi1173_params}}
\tablehead{
\colhead{Parameter} & \colhead{Prior} & \colhead{Posterior}
}
\startdata
$P$ (days) & $\mathcal{U}(7.05, 7.10)$ & $7.0643971 \pm 0.0000017$\\
$T_c$ ($\bjd - 2457000$) & $\mathcal{U}(2529,2530)$ & $2529.37849 \pm 0.00014$\\
$K$ (\ms) & $\mathcal{U}(0, 100)$ & $10.05^{+0.21}_{-0.19}$\\
\secosw & $\mathcal{U}(-1, 1)$ & $-0.053^{+0.053}_{-0.065}$\\
\sesinw & $\mathcal{U}(-1, 1)$ & $0.033^{+0.099}_{-0.094}$\\
\ecosw & Derived & $-0.0055 \pm 0.0076$\\
\esinw & Derived & $0.002^{+0.014}_{-0.011}$\\
$e$ & Derived\tablenotemark{a} & $<0.037$\\
$\dot{\gamma}$ (\msyr) & $\mathcal{U}(-100, 100)$ & $0.4^{+1.2}_{-1.3}$\\
$R_p/R_\star$ & $\mathcal{U}(0.01, 0.5)$ & $0.0882^{+0.0017}_{-0.0018}$\\
$b$ & $\mathcal{U}(0, 1)$ & $0.739^{+0.023}_{-0.016}$\\
$\rho_\star$ (\gcc) & $\mathcal{U}(1.0, 5.0)$ & $1.80^{+0.13}_{-0.17}$\\
$R_p$ (\Rjup) & Derived & $0.783^{+0.025}_{-0.023}$\\
$M_p$ (\Mjup) & Derived & $0.0890^{+0.0033}_{-0.0032}$\\
$\rho_p$ (\gcc) & Derived & $0.230^{+0.019}_{-0.026}$\\
$u_{0,\text{TESS}}$ & $\mathcal{U}(0, 1)$\tablenotemark{b} & $0.49^{+0.34}_{-0.29}$\\
$u_{1,\text{TESS}}$ & $\mathcal{U}(0, 1)$\tablenotemark{b} & $0.14^{+0.37}_{-0.47}$\\
$\gamma_{\text{HIRES}}$ (\ms) & $\mathcal{U}(-100, 100)$ & $-0.2^{+4.5}_{-4.8}$\\
$\gamma_{\text{MX-Blue-1}}$ (\ms) & $\mathcal{U}(-100, 100)$ & $-2.4^{+4.0}_{-3.4}$\\
$\gamma_{\text{MX-Red-1}}$ (\ms) & $\mathcal{U}(-100, 100)$ & $-3.3 \pm 3.8$\\
$\gamma_{\text{MX-Blue-2}}$ (\ms) & $\mathcal{U}(-100, 100)$ & $-2.4^{+3.5}_{-3.4}$\\
$\gamma_{\text{MX-Red-2}}$ (\ms) & $\mathcal{U}(-100, 100)$ & $0.9^{+3.3}_{-3.9}$\\
$\gamma_{\text{MX-Blue-3}}$ (\ms) & $\mathcal{U}(-100, 100)$ & $-1.70^{+0.48}_{-0.47}$\\
$\gamma_{\text{MX-Red-3}}$ (\ms) & $\mathcal{U}(-100, 100)$ & $-1.88^{+0.80}_{-0.82}$\\
$\gamma_{\text{MX-Blue-4}}$ (\ms) & $\mathcal{U}(-100, 100)$ & $-2.07^{+0.25}_{-0.22}$\\
$\gamma_{\text{MX-Red-4}}$ (\ms) & $\mathcal{U}(-100, 100)$ & $-2.18^{+0.25}_{-0.26}$\\
$\log{\sigma_{J,\text{HIRES}}}$ & $\mathcal{U}(-10, 5)$ & $1.25^{+0.22}_{-0.23}$\\
$\log{\sigma_{J,\text{MX-Blue-1}}}$ & $\mathcal{U}(-10, 5)$ & $-5.0^{+2.2}_{-4.6}$\\
$\log{\sigma_{J,\text{MX-Red-1}}}$ & $\mathcal{U}(-10, 5)$ & $-4.6^{+3.8}_{-3.5}$\\
$\log{\sigma_{J,\text{MX-Blue-2}}}$ & $\mathcal{U}(-10, 5)$ & $-4.0^{+4.7}_{-3.0}$\\
$\log{\sigma_{J,\text{MX-Red-2}}}$ & $\mathcal{U}(-10, 5)$ & $-4.4^{+3.9}_{-3.6}$\\
$\log{\sigma_{J,\text{MX-Blue-3}}}$ & $\mathcal{U}(-10, 5)$ & $-5.2^{+2.7}_{-3.8}$\\
$\log{\sigma_{J,\text{MX-Red-3}}}$ & $\mathcal{U}(-10, 5)$ & $-4.8^{+3.6}_{-3.4}$\\
$\log{\sigma_{J,\text{MX-Blue-4}}}$ & $\mathcal{U}(-10, 5)$ & $-0.16^{+0.24}_{-0.21}$\\
$\log{\sigma_{J,\text{MX-Red-4}}}$ & $\mathcal{U}(-10, 5)$ & $-1.6^{+2.1}_{-2.9}$\\
$F_{0,\text{S14}}$ & $\mathcal{N}(0.0,0.01)$ & $0.0036^{+0.0098}_{-0.0094}$\\
$F_{0,\text{S15}}$ & $\mathcal{N}(0.0,0.01)$ & $-0.0015 \pm 0.0099$\\
$F_{0,\text{S21}}$ & $\mathcal{N}(0.0,0.01)$ & $0.0004 \pm 0.0096$\\
$F_{0,\text{S22}}$ & $\mathcal{N}(0.0,0.01)$ & $-0.002 \pm 0.010$\\
$F_{0,\text{S41}}$ & $\mathcal{N}(0.0,0.01)$ & $0.007^{+0.009}_{-0.010}$\\
$F_{0,\text{S47}}$ & $\mathcal{N}(0.0,0.01)$ & $0.002^{+0.009}_{-0.010}$\\
$F_{0,\text{S48}}$ & $\mathcal{N}(0.0,0.01)$ & $0.0065 \pm 0.0096$\\
$F_{0,\text{S74}}$ & $\mathcal{N}(0.0,0.01)$ & $-0.000^{+0.011}_{-0.009}$\\
$F_{0,\text{S75}}$ & $\mathcal{N}(0.0,0.01)$ & $-0.004^{+0.009}_{-0.010}$
\enddata
\tablecomments{$\mathcal{U}(a, b)$ refers to a uniform prior distribution bounded between $a$ and $b$, while $\mathcal{N}(\mu, \sigma)$ refers to a Gaussian distribution centered at $\mu$ and with standard deviation $\sigma$. We report in the posterior column the median value as well as the limits of the 68\% highest-density interval of the posterior distribution.}
\tablenotemark{a}{95\% upper limit.}
\tablenotetext{b}{Uniform in $q_1$, $q_2$ following the parameterization from \citet{Exoplanet_Kipping13}.}
\end{deluxetable}

\begin{deluxetable}{lcc}
\tablecaption{WASP-107 Fitted and Derived Parameters \label{tab:wasp107_params}}
\tablehead{
\colhead{Parameter} & \colhead{Prior} & \colhead{Posterior}
}
\startdata
$P$ (days) & $\mathcal{N}(5.7214742,0.0000043)$\tablenotemark{a} & $5.72148750 \pm 0.00000030$\\
$T_c$ ($\bjd - 2457000$) & $\mathcal{N}(584.329897,0.000032)$\tablenotemark{a} & $584.329899 \pm 0.000031$\\
$K$ (\ms) & $\mathcal{U}(0, 100)$ & $15.38^{+0.46}_{-0.45}$\\
\secosw & $\mathcal{U}(-1, 1)$ & $0.077^{+0.093}_{-0.066}$\\
\sesinw & $\mathcal{U}(-1, 1)$ & $0.02^{+0.12}_{-0.11}$\\
\ecosw & Derived & $0.010^{+0.012}_{-0.014}$\\
\esinw & Derived & $0.001 \pm 0.017$\\
$e$ & Derived\tablenotemark{b} & $<0.052$\\
$R_p/R_\star$ & $\mathcal{U}(0.01, 0.5)$ & $0.1462^{+0.0015}_{-0.0017}$\\
$b$ & $\mathcal{U}(0, 1)$ & $0.21^{+0.13}_{-0.10}$\\
$\log{P_c}$ (days) & $\mathcal{U}(6.9, 7.1)$ & $6.995^{+0.013}_{-0.012}$\\
$P_c$ (days) & Derived & $1091^{+15}_{-13}$\\
$T_{c,c}$ ($\bjdtdb - 2457000$) & $\mathcal{U}(1220,1820)$ & $1543^{+61}_{-55}$\\
$K_c$ (\ms) & $\mathcal{U}(0, 100)$ & $9.5^{+0.9}_{-1.0}$\\
$\secosw_c$ & $\mathcal{U}(-1, 1)$ & $-0.34^{+0.13}_{-0.20}$\\
$\sesinw_c$ & $\mathcal{U}(-1, 1)$ & $-0.43^{+0.10}_{-0.12}$\\
$e_c$ & Derived & $0.315^{+0.075}_{-0.080}$\\
$\rho_\star$ (\gcc) & $\mathcal{U}(1.0, 5.0)$ & $3.39^{+0.26}_{-0.27}$\\
$R_p$ (\Rjup) & Derived & $0.935 \pm 0.023$\\
$M_p$ (\Mjup) & Derived & $0.1039^{+0.0039}_{-0.0041}$\\
$\rho_p$ (\gcc) & Derived & $0.157^{+0.012}_{-0.014}$\\
$u_{0,\text{TESS}}$ & $\mathcal{U}(0, 1)$\tablenotemark{c} & $0.511^{+0.080}_{-0.076}$\\
$u_{1,\text{TESS}}$ & $\mathcal{U}(0, 1)$\tablenotemark{c} & $-0.03^{+0.14}_{-0.22}$\\
$\gamma_{\text{CORALIE}}$ (\ms) & $\mathcal{U}(-100, 100)$ & $1.5^{+1.8}_{-1.6}$\\
$\gamma_{\text{HIRES}}$ (\ms) & $\mathcal{U}(-100, 100)$ & $0.88^{+0.70}_{-0.72}$\\
$\gamma_{\text{MX-Blue-1}}$ (\ms) & $\mathcal{U}(-100, 100)$ & $-0.8^{+1.6}_{-2.0}$\\
$\gamma_{\text{MX-Red-1}}$ (\ms) & $\mathcal{U}(-100, 100)$ & $-0.9^{+1.7}_{-2.0}$\\
$\gamma_{\text{MX-Blue-2}}$ (\ms) & $\mathcal{U}(-100, 100)$ & $-0.5^{+1.9}_{-2.0}$\\
$\gamma_{\text{MX-Red-2}}$ (\ms) & $\mathcal{U}(-100, 100)$ & $-0.5^{+1.8}_{-2.1}$\\
$\log{\sigma_{J,\text{CORALIE}}}$ & $\mathcal{U}(-10, 5)$ & $-3.0^{+5.2}_{-2.7}$\\
$\log{\sigma_{J,\text{HIRES}}}$ & $\mathcal{U}(-10, 5)$ & $1.40^{+0.12}_{-0.11}$\\
$\log{\sigma_{J,\text{MX-Blue-1}}}$ & $\mathcal{U}(-10, 5)$ & $-2.4^{+4.0}_{-3.0}$\\
$\log{\sigma_{J,\text{MX-Red-1}}}$ & $\mathcal{U}(-10, 5)$ & $-4.5^{+2.0}_{-5.5}$\\
$\log{\sigma_{J,\text{MX-Blue-2}}}$ & $\mathcal{U}(-10, 5)$ & $0.93^{+0.21}_{-0.22}$\\
$\log{\sigma_{J,\text{MX-Red-2}}}$ & $\mathcal{U}(-10, 5)$ & $0.84 \pm 0.26$\\
$F_{0,\text{S10}}$ & $\mathcal{N}(0.0,0.01)$ & $0.0071^{+0.0099}_{-0.0097}$\\
$F_{0,\text{S91}}$ & $\mathcal{N}(0.0,0.01)$ & $-0.001^{+0.010}_{-0.009}$
\enddata
\tablenotemark{a}{Ephemeris derived from K2 data by \citet{Dai2017}.}
\tablenotemark{b}{95\% upper limit.}
\tablenotetext{c}{Uniform in $q_1$, $q_2$ following the parameterization from \citet{Exoplanet_Kipping13}.}
\end{deluxetable}

\begin{deluxetable}{lcc}
\tablecaption{HAT-P-18 Fitted and Derived Parameters \label{tab:hatp18_params}}
\tablehead{
\colhead{Parameter} & \colhead{Prior} & \colhead{Posterior}
}
\startdata
$P$ (days) & $\mathcal{N}(5.508023, 0.000006)$\tablenotemark{a} & $5.50802910 \pm 0.00000030$\\
$T_c$ ($\bjd - 2457000$) & $\mathcal{N}(-2284.97751,0.00020)$\tablenotemark{a} & $-2284.97750^{+0.00019}_{-0.00020}$\\
$K$ (\ms) & $\mathcal{U}(0, 100)$ & $23.97 \pm 0.40$\\
\secosw & $\mathcal{U}(-1, 1)$ & $-0.046^{+0.057}_{-0.061}$\\
\sesinw & $\mathcal{U}(-1, 1)$ & $0.029 \pm 0.088$\\
\ecosw & Derived & $-0.0041^{+0.0070}_{-0.0067}$\\
\esinw & Derived & $0.002^{+0.011}_{-0.009}$\\
$e$ & Derived\tablenotemark{b} & $<0.03$\\
$\dot{\gamma}$ (\msyr) & $\mathcal{U}(-100, 100)$ & $-2.2^{+1.5}_{-1.6}$\\
$R_p/R_\star$ & $\mathcal{U}(0.01, 0.5)$ & $0.1344^{+0.0023}_{-0.0030}$\\
$b$ & $\mathcal{U}(0, 1)$ & $0.23^{+0.18}_{-0.13}$\\
$\rho_\star$ (\gcc) & $\mathcal{U}(1.0, 5.0)$ & $3.22^{+0.33}_{-0.25}$\\
$R_p$ (\Rjup) & Derived & $0.955^{+0.032}_{-0.030}$\\
$M_p$ (\Mjup) & Derived & $0.1747^{+0.0053}_{-0.0052}$\\
$\rho_p$ (\gcc) & Derived & $0.249^{+0.024}_{-0.026}$\\
$u_{0,\text{TESS}}$ & $\mathcal{U}(0, 1)$\tablenotemark{c} & $0.44^{+0.11}_{-0.14}$\\
$u_{1,\text{TESS}}$ & $\mathcal{U}(0, 1)$\tablenotemark{c} & $0.22 \pm 0.30$\\
$\gamma_{\text{HIRES}}$ (\ms) & $\mathcal{U}(-100, 100)$ & $-21 \pm 10$\\
$\gamma_{\text{HARPS-N}}$ (\ms) & $\mathcal{U}(-100, 100)$\tablenotemark{d} & $-5.5^{+2.5}_{-2.6}$\\
$\gamma_{\text{MX-Blue-1}}$ (\ms) & $\mathcal{U}(-100, 100)$ & $24 \pm 16$\\
$\gamma_{\text{MX-Red-1}}$ (\ms) & $\mathcal{U}(-100, 100)$ & $24 \pm 16$\\
$\gamma_{\text{MX-Blue-2}}$ (\ms) & $\mathcal{U}(-100, 100)$ & $24^{+17}_{-15}$\\
$\gamma_{\text{MX-Red-2}}$ (\ms) & $\mathcal{U}(-100, 100)$ & $27^{+15}_{-17}$\\
$\log{\sigma_{J,\text{HIRES}}}$ & $\mathcal{U}(-10, 5)$ & $2.79^{+0.13}_{-0.14}$\\
$\log{\sigma_{J,\text{HARPS-N}}}$ & $\mathcal{U}(-10, 5)$ & $-4.3^{+4.0}_{-3.6}$\\
$\log{\sigma_{J,\text{MX-Blue-1}}}$ & $\mathcal{U}(-10, 5)$ & $-0.1^{+1.3}_{-2.3}$\\
$\log{\sigma_{J,\text{MX-Red-1}}}$ & $\mathcal{U}(-10, 5)$ & $-2.6^{+4.0}_{-2.8}$\\
$\log{\sigma_{J,\text{MX-Blue-2}}}$ & $\mathcal{U}(-10, 5)$ & $-5.0^{+1.8}_{-5.0}$\\
$\log{\sigma_{J,\text{MX-Red-2}}}$ & $\mathcal{U}(-10, 5)$ & $-4.0^{+5.2}_{-2.5}$\\
$F_{0,\text{S25}}$ & $\mathcal{N}(0.0,0.01)$ & $-0.002^{+0.009}_{-0.010}$\\
$F_{0,\text{S26}}$ & $\mathcal{N}(0.0,0.01)$ & $0.001^{+0.009}_{-0.010}$\\
$F_{0,\text{S79}}$ & $\mathcal{N}(0.0,0.01)$ & $0.000 \pm 0.010$
\enddata
\tablenotetext{a}{Ephemeris from \citet{Hartman2011}.}
\tablenotemark{b}{95\% upper limit.}
\tablenotetext{c}{Uniform in $q_1$, $q_2$ following the parameterization from \citet{Exoplanet_Kipping13}.}
\tablenotetext{d}{A fixed offset of $-11100~\ms$ was removed from the HARPS-N data before fitting.}
\end{deluxetable}

\onecolumngrid
\section{Robustness Checks \label{sec:robustness}}

In this section, we assess the robustness of our eccentricity constraints.
While fitting the RV data, we allowed for a per-instrument ``jitter'' term $\sigma_J$ to account for unmodeled white noise arising from either instrumental or astrophysical sources.
For our observations of TOI-1173 and HAT-P-18, all best-fit jitter values for the most precise MAROON-X data were $\sigma_J < 1.0\,\ms$, below the per-point instrumental uncertainties.
For WASP-107, the second observing run of MAROON-X data required $\sigma_J \sim 2.5\,\ms$, suggesting at an additional source of noise.
WASP-107 is known to be a moderately spotted star \citep{Dai2017}, which may give rise to spurious RV variation due to rotation of spots in and out of the field of view.
We performed a Lomb-Scargle periodogram analysis of the RV residuals after subtracting the best-fit model for the two planets in the WASP-107 system and found no significant peaks, including at the 17 day stellar rotation period.
We also checked that there were no correlations between the RV residuals and the differential line width or other stellar activity indicators.
We therefore conclude that our eccentricity constraints are unlikely to be biased by the presence of time-correlated or activity-induced RV signals.

Still, all eccentricity measurements from RV data are subject to the \citet{LucySweeney71} bias because $e$ is a strictly positive quantity.
Since the 68\% highest-density interval of the eccentricity posterior probability distribution contains $e = 0$, we interpret our constraints as purely upper limits rather than a measurement of a non-zero eccentricity.
Any unrecognized bias would therefore only strengthen our conclusions that these planets are not being actively heated by eccentricity tides.
\citet{Hara2019} used a suite of numerical simulations to estimate the expected bias in eccentricity inferred from an RV dataset, finding that if the true eccentricity is zero,
\begin{equation}
    b_e = \frac{\sigma_\mathrm{RV}}{K}\sqrt{\frac{\pi}{N-p}},
\end{equation}
where $N-p$ is the number of data points less the number of fitted free parameters.
Using the mean instrumental uncertainty, we find that the expected bias in our eccentricity measurement is $b_e = 0.023$ for TOI-1173\,b, $b_e = 0.019$ for WASP-107\,b, and $b_e = 0.027$ for HAT-P-18\,b, comparable to the constraints we achieved with the new data.

\begin{figure}
    \includegraphics[width=0.32\textwidth]{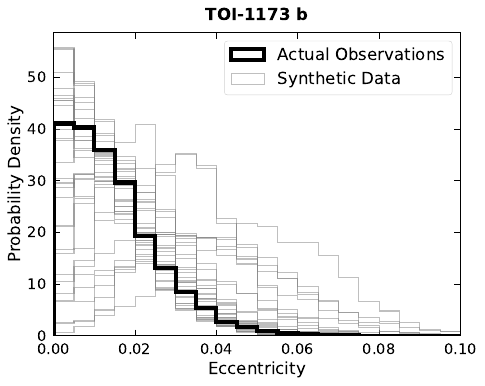}
    \includegraphics[width=0.32\textwidth]{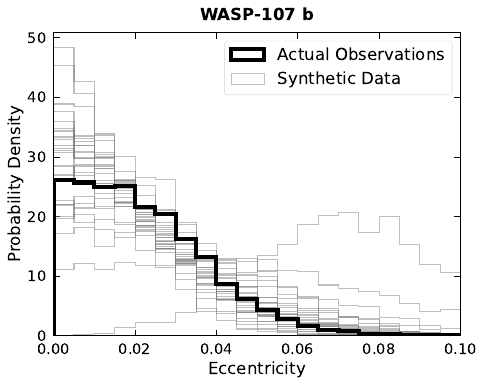}
    \includegraphics[width=0.32\textwidth]{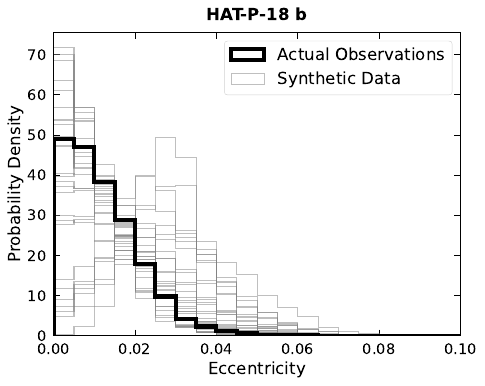}
    \caption{The eccentricity posterior probability distribution for actual observed data (dark line) and 30 simulated datasets assuming circular orbits (light gray lines). These simulations indicate that our constraints are in line with that expected given circular orbits and the quality of our data.}
    \label{fig:synth_ecc_hist}
\end{figure}

We verified these bias estimates by performing simulations tailored to our specific dataset, generating synthetic data with the actual observed timestamps and uncertainties, but arising from a purely circular orbit.
Figure \ref{fig:synth_ecc_hist} shows the eccentricity posteriors for the observed data and 30 simulated datasets.
We find that the median 68\% upper limit on $e$ that we were able to recover from the synthetic data was $e < 0.02$ for the two planets.
We therefore confirm the analytic estimates from \citet{Hara2019} and demonstrate that the observed data are fully consistent with circular orbits for all three planets.

\bibliography{main,software,instruments}{}
\bibliographystyle{aasjournalv7}

\end{document}